\documentclass[12pt]{article}

\usepackage{bm}
\usepackage{amsmath}
\usepackage[bookmarks, colorlinks=true, plainpages = false, 
citecolor = green, urlcolor = blue, filecolor = blue] {hyperref}
\usepackage[font=small,labelfont=bf,textfont=it]{caption}

\usepackage{graphicx}

\begin{document}

\title{Understanding Quaternions and the Dirac Belt Trick}         
\author{Mark Staley\footnote{Adjunct Professor, University of Ontario Institute of Technology (now Ontario Tech University). Present address: King's Hill Research, email: \protect\href{mailto:mstaley@kingshillresearch.com}{mstaley@kingshillresearch.com}. This is the Accepted Manuscript version of an article accepted for publication in European Journal of Physics. IOP Publishing Ltd is not responsible for any errors or omissions in this version of the manuscript or any version derived from it. The Version of Record is available online at \protect\url{https://doi.org/10.1088/0143-0807/31/3/004}.}}        
\date{February 14, 2010}          
\maketitle

\begin{abstract}
The Dirac belt trick is often employed in physics classrooms to show that a $2\pi$ rotation is not topologically equivalent to the absence of rotation whereas a $4\pi$ rotation is, mirroring a key property of quaternions and their isomorphic cousins, spinors.  The belt trick can leave the student wondering if a real understanding of quaternions and spinors has been achieved, or if the trick is just an amusing analogy.  The goal of this paper is to demystify the belt trick and to show that it suggests an underlying \emph{four-dimensional} parameter space for rotations that is simply connected.  An investigation into the geometry of this four-dimensional space leads directly to the system of quaternions, and to an interpretation of three-dimensional vectors as the generators of rotations in this larger four-dimensional world.  The paper also shows why quaternions are the natural extension of complex numbers to four dimensions.  The level of the paper is suitable for undergraduate students of physics.   
\end{abstract}

\newpage

\section{Introduction} 
Every student learns that complex numbers are necessary to solve certain algebraic equations such as $x^2+1=0$.  The student also learns that complex numbers may be represented as two-dimensional vectors lying in the so-called Argand plane, with the x-axis representing the real numbers and the y-axis representing the pure-imaginary numbers.  The geometric interpretation of complex numbers is supported by the observation that the addition of two complex numbers represents the addition of two vectors, and the product of two unit-length complex numbers represents a sequence of two rotations.  

Upon seeing the connection between complex numbers and two-dimensional geometry, the curious mind is apt to wonder whether there is an extension applicable to three-dimensional geometry.  During the nineteenth century, this germ of an idea was lodged in the minds of many people including William Rowan Hamilton.  In 1843 Hamilton discovered an algebra of three-dimensional rotations that was based on a new set of objects he called \emph{quaternions} (van der Waerden, 1976 provides a good history).  So impressed was he with these new objects that he spent the remaining 22 years of his life working out their properties (Crowe, 1967).  After Hamilton died his quaternions were mostly abandoned, leading the biographer E.T. Bell to suggest that the great man had been the victim of a mono-maniacal delusion (Bell, 1937).  

Developments of the twentieth century have largely vindicated Hamilton.  Firstly, historians have shown that Gibbs and Heaviside were heavily influenced by Hamilton's work when they developed modern vector analysis (Stephenson, 1966, Crowe, 1967, Silva and Martins, 2002). Secondly, quaternions are now widely used in the computer graphics and aerospace industries to reduce the computational costs associated with performing rotations on vectors (Kuipers, 1999).  And thirdly \emph{spinors}, those fundamental building blocks of modern particle physics, are isomorphic to quaternions (Kronsbein, 1967).  On this third point it is interesting to note that Hamilton's nineteenth century belief that quaternions held the key to understanding the universe has its modern parallel in the view of some physicists that spinors are more fundamental than space-time vectors (Bohm, 1971, Penrose \& Rindler, 1986).               

While spinors and quaternions are undoubtedly important and useful, their geometrical properties can be difficult to understand.\footnote{Girard (1984) reviews various applications of quaternions in physics.  Coddens (2002) discusses the difficulty faced by students in understanding spinor concepts.}  Take for example the Dirac belt trick.  The trick is motivated by the observation that when a spinor (such as an electron) is rotated by $2\pi$ its quantum mechanical wave function reverses sign, which has observable implications (Silverman, 1980).  A second rotation of $2\pi$ restores the wave function back to its original form.  Dirac came up with the belt trick as a way to demonstrate this effect using an everyday three-dimensional object, and it is often used in undergraduate physics classrooms.  The trick shows that a $2\pi$ rotation is not equivalent to no rotation although a $4\pi$ rotation is.  A belt is held fixed at one end while the other end is rotated (twisted) through an angle of $4\pi$ about an axis parallel to its length.  It is shown that the belt can be untwisted without any further rotation by simply passing one end of the belt under the rest of the belt (see Section 3).  Hence a $4\pi$ rotation is topologically equivalent to no rotation.             

The belt trick can leave the student wondering if a real understanding of quaternions and spinors has been achieved, or if the trick is just an amusing analogy.  And deeper study of quaternions and spinors leads to further questions, such as why do they have four components (or two complex components), whereas ordinary spatial geometry gets by with three?  Or why do rotations in three-dimensional space become similarity transformations when expressed in the language of quaternions?  There are several approaches to answering these questions, the most common of which is based on group theory.  One can define spinor transformations using the SU(2) group and show that these are isomorphic to (and form a double cover of) SO(3), the three-dimensional rotation group (see for example Joshi, 1982).  Other approaches use geometrical pictures to aid in understanding.  For example there is an approach based on projective geometry (Bohm, 1971, Frescura and Hiley, 1981), and another approach based on the ``vector plus flag picture'' (Misner, Thorne and Wheeler, 1973).  

This paper presents a geometrical approach to understanding complex numbers and quaternions that is based on equating their elements with generators of rotations.  The unit imaginary ``i'' in complex number theory can be interpreted as the generator of rotations in two dimensions.  In four dimensions there are six generators of rotations, and these can be decomposed into two sets of generators each consisting of three elements corresponding to the hyper-complex numbers in Hamilton's system of quaternions.  This paper will show that each of these triplets can be interpreted as the basis of a coordinate system in three dimensions, similar to the way that a point in the Argand plane can be used to represent a vector in two dimensions.  One of these triplets forms the basis of a right-handed coordinate system while the other forms the basis of a left-handed coordinate system.  

The paper is organized as follows.  Section 2 presents an interpretation of complex numbers in terms of two-dimensional rotations that allows for an extension to higher dimension.  Section 3 discusses three dimensional geometry and the Dirac belt trick.  The purpose of this section is to demystify the belt trick and to motivate the development of quaternions.  Section 4 describes the algebraic and geometrical properties of quaternions and shows how ordinary three-dimensional geometry is reclaimed.  Finally, Section 5 concludes.

\section{Two-Dimensional Geometry} 
Consider the rotation of a vector in two dimensions:
\[
\bm {v'} = R \bm v,
\]
where 
\[
R = \left(\begin{array}{cc} \cos\theta & -\sin\theta  \\ \sin\theta & \cos\theta \end{array}\right), \quad
\bm {v} = \left(\begin{array}{c} x  \\ y     \end{array}\right)
\]
Now re-write $R$ as 
\[
R=I  \cos \theta  + i  \sin  \theta ,
\] 
where
\[
I = \left(\begin{array}{cc} 1 & 0  \\ 0 & 1 \end{array}\right), 
\quad i = \left(\begin{array}{cc} 0 & -1  \\ 1 & 0 \end{array}\right)
\]
The following multiplication table may be used to combine rotation operations: 
\begin{equation}\label{1}
\begin{array}{c|cc}
     & I & i \\
  \hline
     I & I & i \\
     i & i & -I \\
\end{array} 
\end{equation} 
The above algebra is isomorphic to the system of complex numbers, with $I$ standing for $1$ and $i$ standing for the unit imaginary ``$\sqrt{-1}$''.  Hence rotations can be represented using complex numbers.

Recall that a rotation can be constructed out of a large sequence of infinitesimal rotations using a so-called \emph{generator}.  We can now show that the symbol $i$ is the generator of rotations in two dimensions.  Let's say we want to apply two consecutive rotations to a vector.  Call the first rotation $R_1=\cos\theta_1\,I+\sin\theta_1\,i$ and the second $R_2=\cos\theta_2\,I+\sin\theta_2\,i$.  Using Table \eqref{1} and some simple trigonometric identities we can show that $R_1 R_2=\cos(\theta_1+\theta_2)I+\sin(\theta_1+\theta_2)i$.  So to combine rotations we simply add the angles together.  By extension, a rotation by a given angle $\theta$ is equivalent to a sequence of $N$ rotations each of angle $\theta/N$.  Taking the limit as $N \to\infty$ we have
\begin{equation}\label{generator}
R=I \, \cos \theta  + i \, \sin  \theta  = \mathop {\lim }\limits_{N\to\infty} \left( I+i\,\frac{\theta}{N} \right) ^N,
\end{equation}        
where we have used the property that $\cos(\alpha) \to 1$ as $\alpha \to 0$ and $\sin(\alpha) \to \alpha$ as $\alpha \to 0$.  The quantity inside the brackets on the right hand side of Equation \eqref{generator} is of the form of an infinitesimal rotation, and the generator is $i$.

Let us now investigate the parameter space of two-dimensional rotations.  Figure 1 shows one way to parameterize rotations using a single dimension corresponding to the angle $\theta$.  We may note that the angles $\pi$ and $-\pi$ are equivalent in this representation.  Figure 2 shows how to construct an alternative parameter space by increasing the the number of dimensions to two using $I$ and $i$.  The coordinates of a point representing a rotation are constrained to lie on the unit circle.     

\begin{figure}[htbp] 
  \centering
\vspace{.3in}

  \includegraphics[width=4.42in,keepaspectratio]{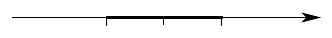}
  \put(-105,0){\fontsize{14.23}{17.07}\selectfont $\pi$}
  \put(-158,0){\fontsize{14.23}{17.07}\selectfont $0$}
  \put(-225,0){\fontsize{14.23}{17.07}\selectfont $-\pi$}
  \put(-5,12){\fontsize{14.23}{17.07}\selectfont $\theta$}
  \caption{Parameter Space for 2-d Rotations}
  \vspace{.3in}
  \label{fig:1dLine}
\end{figure}

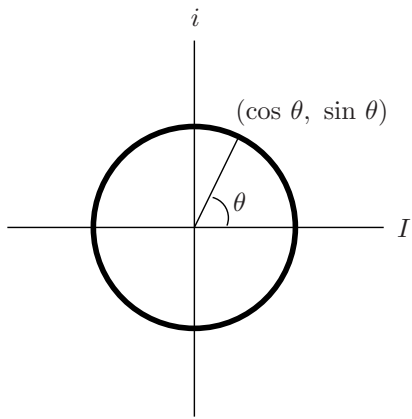
\begin{figure}[htbp] 
  \centering
  \includegraphics[width=2.18in,keepaspectratio]{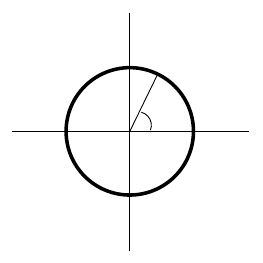}
  \put(-65,85){\fontsize{10.23}{17.07}\selectfont $\theta$}
  \put(-3,75){\fontsize{10.23}{17.07}\selectfont $I$}
  \put(-81,155){\fontsize{10.23}{17.07}\selectfont $i$}
  \put(-64,120){\fontsize{10.23}{17.07}\selectfont $(\cos\,\theta,\,\,\sin\,\theta )$}
  \caption{Alternative Parameter Space for 2-d Rotations}
  \vspace{.3in}
  \label{fig:2dSpace}
\end{figure}

We can introduce scaling operations into our algebra by admitting complex numbers having non-unit magnitudes.  A general operation consisting of a rotation and a scaling operation can be written 
\[
R=r \{I \cos \theta  + i \sin  \theta \}, \quad r>0,
\]
which can be represented in Figure 2 by a point lying anywhere in the plane (i.e. it is no longer constrained to lie on the unit circle).  We can interpret the plane in Figure 2 as the \emph{Argand plane}.  

A point in the Argand plane can be used to represent a rotation/scaling operation, but it can also be used to represent a two-dimensional vector.  This duality can be made clearer by noting that the unit basis vectors (in column format) are related to $I$ and $i$ as follows
\[
\left(\begin{array}{c} 1 \\ 0 \end{array}\right) = I \left(\begin{array}{c} 1 \\ 0 \end{array}\right) \quad \text{and} \quad \left(\begin{array}{c} 0 \\ 1 \end{array}\right) = i \left(\begin{array}{c} 1 \\ 0 \end{array}\right)
\]
which means that an arbitrary vector $(a,b)$ (here in row format) can be obtained by rotating and scaling the the unit vector $(1,0)$ using $aI+bi$.  This implies that we can rotate $(a,b)$ by the angle $\theta$  and scale it by a factor of $r$ by simply multiplying $aI+bi$ by $r \{\cos \theta \, I  +  \sin \theta \, i \}$ and reading off the components of the resulting complex number.  Hence we can dispense with vectors altogether and use complex numbers to both represent vectors and rotation/scaling operations.  Note that for pure rotations the parameter ``$r$'' is equal to one.  We will see later that a similar formalism holds for quaternions.    

The goal of the following two sections will be to extend the above reasoning to three and four dimensions.  In doing so we will endeavor to create something like a multi-dimensional extension of the Argand plane.  That is, we will wish to build a parameter space that can act as a Euclidean vector space.  One useful property of Euclidean spaces that can be used to test if a parameter space is admissible for such purposes is that it be \emph{simply connected}.  A space is simply-connected if one can continuously deform any path connecting two end-points into any other path connecting the same two end-points without moving the end-points themselves.  A special case is where the starting and ending points are the same, in which case the path is a loop.  If one can shrink the loop down to a point without moving the starting/ending points, then the space is simply connected.  The Argand plane is an example of a simply-connected space.

\section{Three-Dimensional Rotations}
The rotation of a vector $\bm v$ in three dimensions about an axis of rotation $\bm {u}$ (of unit norm) by an angle $\theta$ is given by
\begin{equation}\label{3}
\bm {v'} = \bm v_\| + \bm v_\bot \cos \theta + (\bm u \times \bm v_\bot ) \sin \theta   
\end{equation}
where $\bm v_\|=\bm u (\bm u \cdot \bm v)$ is the component of $\bm v$ in the direction of $\bm u$, and $\bm v_\bot = \bm v - \bm u (\bm u \cdot \bm v)$ is the component of $\bm v$ perpendicular to $\bm u$ (i.e. in the plane of rotation). 

Consider the parameter space for three-dimensional rotations.  Any given rotation can be represented by a unit vector $\bm u$ and an angle $\theta$ which takes on values between $-\pi$ and $\pi$.  Multiplying the set of unit vectors $\{ \bm u \}$ by the parameter $\theta$ we have a solid sphere of radius $\pi$, which represents the parameter space of rotations.  This parameter space is similar to that shown in Figure 1 in the sense that a rotation about any axis $\bm u$ through an angle of $\pi$ is equivalent to a rotation about that same axis through an angle of $-\pi$.  Recall that in the case of two dimensional rotations it was possible to extend the dimensionality of the parameter space from one to two, and so construct an alternative parameter space in two dimensions that could be mapped in a one-to-one manner with the underlying vector space (see Figure 2).  In the case of three-dimensional rotations our parameter space is already of dimension three, so any attempt to create a similar ``Argand-space'' would require an increase in the number of dimension beyond three.  Before we attempt to go down that road, let us for the moment confine ourselves to three dimensions and continue to explore the properties of the parameter space of rotations as envisioned so far.  

To simplify the discussion even further, let us restrict the dimensionality of the parameter space to two, in which case the parameter space is a \emph{disk} of radius $\pi$, representing rotations about two axes.  The center of the disk corresponds to no rotation and is the starting point for any sequence of rotations.  A rotation of angle $\theta$ about an axis $\bm u$ can be represented in the disk by drawing a line segment of length $\theta$ extending from the center of the disk in the direction $\bm u$.  A rotation of angle $\pi$ is represented by a line segment extending from the center of the disk to the perimeter of the disk (see the examples in Figure 3).    

Another way to study the parameter space of rotations about two axes is to use a belt.  The twists of a belt can serve to keep track of the path followed by a series of rotations.  To see how this is so, lay a belt flat on a table.  You may verify that it is possible to bend or twist the belt about any axis in the plane of the table, but it is impossible to bend it about the vertical axis without ripping it apart.\footnote{In reality you can deform a belt about the vertical axis, but you will see that this deformation is accomplished by a succession of deformations about axes lying in the plane of the table.}  Hence a belt can be twisted about two axis only.  Now hold one end of the belt with your left hand (the end with no buckle) and twist the other end (the end with the buckle) about an axis parallel to the length of the belt.  You will see that the various orientations of the buckle are recorded as varying amounts of twist at different points along the length of the belt.  If you twist the belt by $\pi$ radians, you will see something like the picture in Figure 3(a).  

So we now have two ways of recording the path in parameter space taken by any sequence of rotations about two axes.  Figure 3 illustrates the correspondence between various twists of the belt and points in the parameter disk.  In Figure 3(a) the belt is twisted by an angle $\pi$ about the axis parallel to the length of the belt.  In this case the axis of rotation is in a direction from left to right.  This is represented in the disk as a line segment of length $\pi$ extending left to right from the center of the disk to the perimeter of the disk.  In Figure 3(b) the belt is twisted by an angle $\pi$ about an axis in the plane of the table that is perpendicular to the length of the belt.  In that case the belt is coiled by half a revolution.  This is represented in the disk as a line segment of length $\pi$ extending from the center of the disk up to the top of the perimeter of the disk.

\begin{figure}[tp]
	\centering
	\begin{minipage}[c]{0.45in}\centering (a)\end{minipage}%
	\begin{minipage}[c]{1.4in}\includegraphics[width=\linewidth]{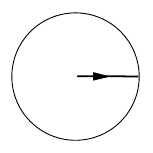}\end{minipage}%
	\hspace{0.2in}%
	\begin{minipage}[c]{1.2in}\includegraphics[width=\linewidth]{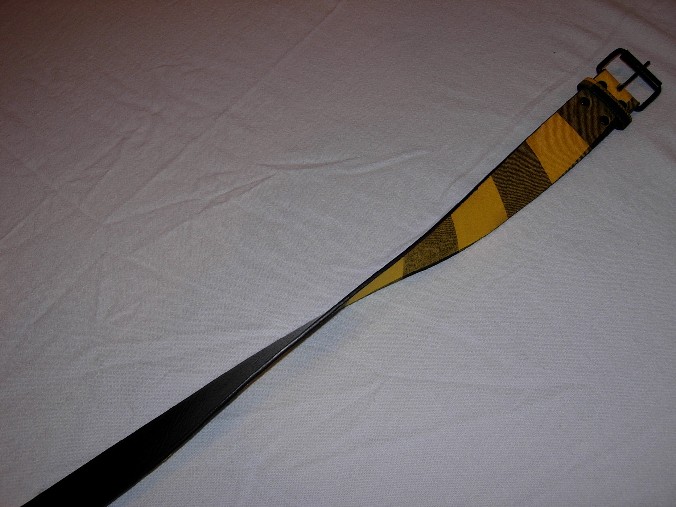}\end{minipage}\\[0.8em]
	
	\begin{minipage}[c]{0.45in}\centering (b)\end{minipage}%
	\begin{minipage}[c]{1.4in}\includegraphics[width=\linewidth]{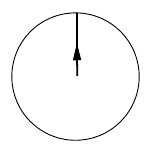}\end{minipage}%
	\hspace{0.2in}%
	\begin{minipage}[c]{1.2in}\includegraphics[width=\linewidth]{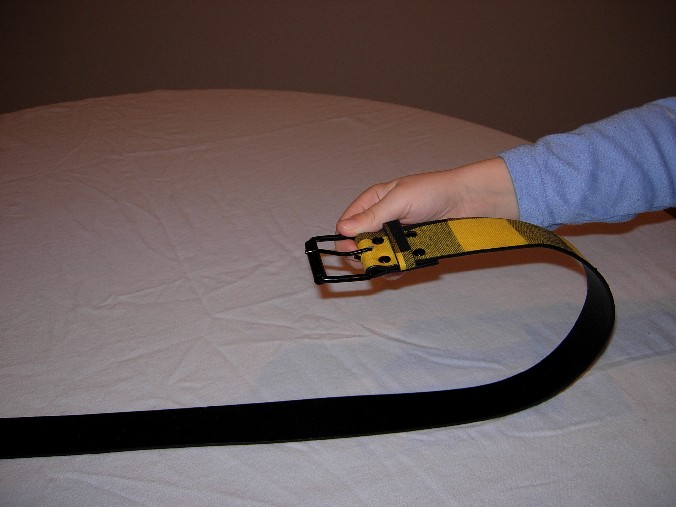}\end{minipage}
	
	\caption{Rotation by $\pi$ about two different axis.  In Figure 3(a) the belt is twisted by an angle $\pi$ about the axis parallel to the length of the belt, which is represented in the disk as a line segment of length $\pi$ extending left to right from the center of the disk to the perimeter of the disk.  In Figure 3(b) the belt is rotated by an angle $\pi$ about an axis in the plane of the table perpendicular to the length of the belt (the belt buckle is held in the air above the table).  This rotation is represented in the disk as a line segment of length $\pi$ extending from the center of the disk up to the top of the perimeter of the disk.}
	\label{fig:HalfTwist}
\end{figure}

\begin{figure}[p]
	\centering
	\begin{minipage}[c]{0.45in}\centering (a)\end{minipage}%
	\begin{minipage}[c]{1.25in}\includegraphics[width=\linewidth]{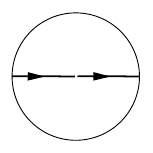}\end{minipage}%
	\hspace{0.2in}%
	\begin{minipage}[c]{1.05in}\includegraphics[width=\linewidth]{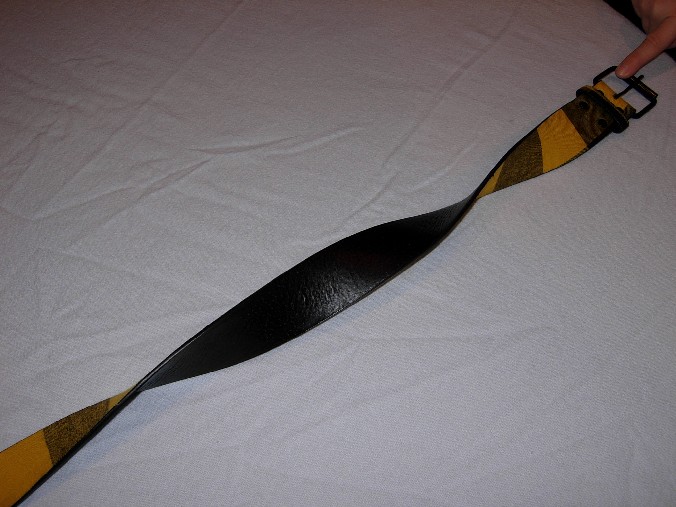}\end{minipage}\\[0.6em]
	
	\begin{minipage}[c]{0.45in}\centering (b)\end{minipage}%
	\begin{minipage}[c]{1.25in}\includegraphics[width=\linewidth]{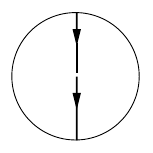}\end{minipage}%
	\hspace{0.2in}%
	\begin{minipage}[c]{1.05in}\includegraphics[width=\linewidth]{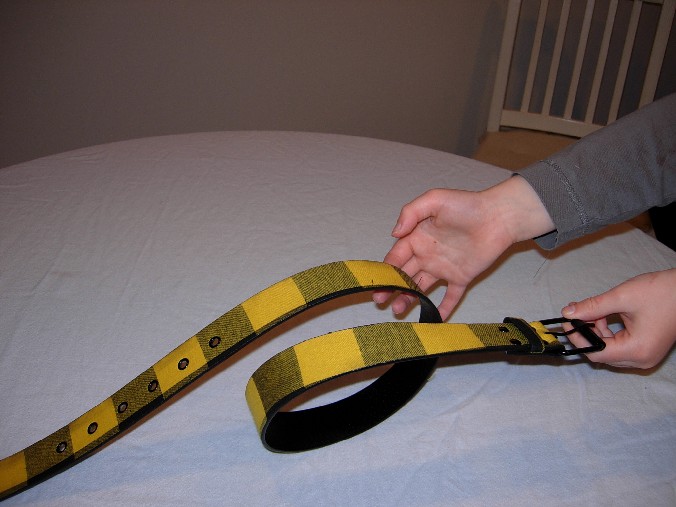}\end{minipage}\\[0.6em]
	
	\begin{minipage}[c]{0.45in}\centering (c)\end{minipage}%
	\begin{minipage}[c]{1.25in}\includegraphics[width=\linewidth]{FullCoilDisk}\end{minipage}%
	\hspace{0.2in}%
	\begin{minipage}[c]{1.05in}\includegraphics[width=\linewidth]{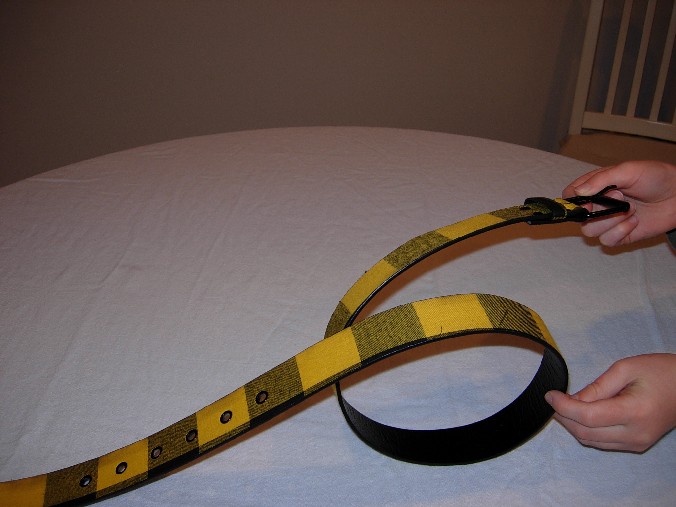}\end{minipage}\\[0.6em]
	
	\begin{minipage}[c]{0.45in}\centering (d)\end{minipage}%
	\begin{minipage}[c]{1.25in}\includegraphics[width=\linewidth]{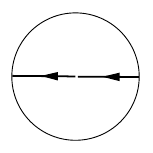}\end{minipage}%
	\hspace{0.2in}%
	\begin{minipage}[c]{1.05in}\includegraphics[width=\linewidth]{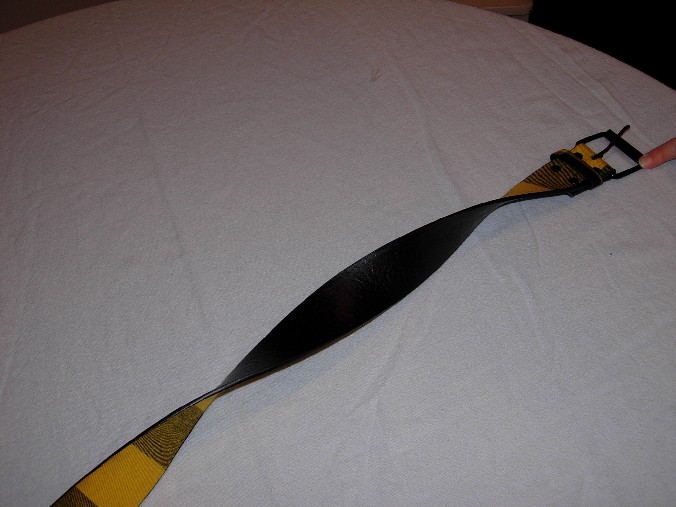}\end{minipage}
	
	\caption{Deformation of $2\pi$ Rotations.  Parameter paths in the disk corresponding to $2\pi$ rotations can be deformed into one another by moving the perimeter points such that they are always located at opposite ends of the disk.  There is no need to move the starting and ending points in the center of the disk.  These deformations correspond to movements of the belt in which the buckle is kept in a fixed orientation.  In the above sequence, a $2\pi$ rotation is deformed into a $-2\pi$ rotation about the same axis.  The transition from (b) to (c) requires that the buckle be passed through the center of the coil.}
	\label{fig:FullTwist}
\end{figure}

Figure \ref{fig:FullTwist} depicts a series of $2\pi$ rotations.  Recall that a rotation of $\pi$ radians about some axis is equivalent to a rotation of $-\pi$ radians around the same axis, which means that any two points located on opposite sides of the parameter disk are equivalent.  The way to represent a $2\pi$ rotation using the parameter disk is to first draw a line from the center of the disk to a point on the perimeter of the disk, then jump to the opposite end of the disk and continue drawing the line back to the center of the disk.  For example, Figure \ref{fig:FullTwist}(a) shows how a $2\pi$ rotation about the axis parallel to the length of the belt is represented in the disk.  Note that the starting and ending points of the parameter path are both represented as points located in the center of the disk, and these correspond to the starting and ending orientations of the belt buckle. 

As shown by the disks in Figure \ref{fig:FullTwist}, parameter paths corresponding to $2\pi$ rotations can be deformed into one another by moving the perimeter points such that they are always located at opposite ends of the disk.  There is no need to move the starting and ending points in the center of the disk.  These deformations correspond to movements of the belt in which \emph{the buckle is kept in a fixed orientation}.  For example, to change the orientation from that shown in Figure \ref{fig:FullTwist}(a) to that of \ref{fig:FullTwist}(b), one need only move the belt buckle towards the left.  The twist then changes into a coil.  The transition from \ref{fig:FullTwist}(b) to \ref{fig:FullTwist}(c) requires that the belt buckle be passed under the belt, i.e. through the center of the coil.  The final move consists of pulling the belt buckle back towards the right.  This sequence of moves shows that a rotation of $2\pi$ about one axis can be continuously deformed into a rotation of $2\pi$ about another axis, or into a rotation of $-2\pi$ about the original axis, without changing the end points.  

An important observation is that it is not possible to deform a belt that has been twisted by $2\pi$ into a flat belt without changing the orientation of the buckle.  Similarly, it is not possible to deform any of the paths shown in the disks in Figure \ref{fig:FullTwist} into a single point in the center of the disk.  This means that our parameter space for rotations is \emph{not} simply connected.  Recalling the discussion at the end of Section 2, our parameter space does not appear to be a good candidate for representing three-dimensional Euclidean vectors.  We cannot construct a three-dimensional ``Argand-space'' out of our disk.

We are now in a position to see why a rotation by $4\pi$ is in some sense equivalent to no rotation.  Twist the belt by $4\pi$ about the axis parallel to the belt.  Then, keeping the first $2\pi$ twist intact, deform the second $2\pi$ twist as shown in Figure \ref{fig:FullTwist}.  The result is a $2\pi$ twist followed by a $-2\pi$ twist, which cancels the first.  Normally this belt trick is done by passing the entire twisted belt under itself, but some experimentation shows that such a maneuver is equivalent to the one just described.  

The lesson from the above exercise is that if we wish to build a simply-connected parameter space for rotations, we should keep in mind that from any given starting point all paths corresponding to a $2\pi$ rotation should correspond to the same end point.  And a rotation of $4\pi$ is equivalent to no rotation, and so can be thought of as a closed loop in a simply-connected parameter space.  There remains the question of whether a $2\pi$ rotation is equivalent to no rotation.  Our every-day experience tells us that the answer to that question is yes, but there remains the possibility that we will have to give up that notion when constructing a simply-connected parameter space.  If that is the case, then a $2\pi$ rotation and a zero $\pi$ rotation will be located at different points within the parameter space, and the mapping between points of the parameter space and points in the underlying Euclidean vector space will not be as as straightforward as was the case in two dimensions.

Recalling our earlier discussion suggesting the need to expand the number of dimensions, we are immediately led to wonder if a sphere could be used to represent rotations about two axis.  Taking the Earth as an example, we could use the north pole as the starting point for rotations, with the direction of each line of longitude representing the direction of the axis of rotation.  The lines of latitude would represent the angles of rotation.  A $2\pi$ rotation might correspond to a line of longitude circling the Earth and ending back at the north pole.  One can immediately see that this scheme does not work.  The lines of longitude all cross each other at the south pole, implying that all rotations by $\pi$ are equivalent to one another, which is obviously not true.  However, a small modification leads to a workable parameter space.  All one need do is equate the south pole with $2\pi$ rotations.   This may be accomplished by setting the actual angle of rotation to be twice the angle depicted by the line of latitude.   Now all $2\pi$ rotations are represented by the same point at the south pole, all paths connecting the north and south poles are continuously deformable into one another (the parameter space is simply connected), and a $4\pi$ rotation (which takes us back to the north pole) is equivalent to no rotation.  Note that a circle around the globe can be shrunk down to a point.  Extending this geometrical insight, we should be able to parameterize three dimensional rotations in a simply-connected manner using a \emph{four-dimensional} hyper-sphere.  So let us turn to a study of four dimensional Euclidean geometry.

\section{Quaternions}
Generalizing from the two dimensional geometry described in Section 2, there are six planes of rotation in four-dimensional space and hence six generators of rotations:
\[
T_{12} = \left(\begin{array}{cccc} 0 & -1 & 0 & 0  \\ 1 & 0 & 0 & 0 \\ 0 & 0 & 0 & 0 \\ 0 & 0 & 0 & 0   \end{array}\right) \quad
T_{23} = \left(\begin{array}{cccc} 0 & 0 & 0 & 0  \\ 0 & 0 & -1 & 0 \\ 0 & 1 & 0 & 0 \\ 0 & 0 & 0 & 0   \end{array}\right)  
\]
\[ 
T_{13} = \left(\begin{array}{cccc} 0 & 0 & -1 & 0  \\ 0 & 0 & 0 & 0 \\ 1 & 0 & 0 & 0 \\ 0 & 0 & 0 & 0   \end{array}\right) \quad
T_{24} = \left(\begin{array}{cccc} 0 & 0 & 0 & 0  \\ 0 & 0 & 0 & -1 \\ 0 & 0 & 0 & 0 \\ 0 & 1 & 0 & 0   \end{array}\right) 
\]
\[  
T_{14} = \left(\begin{array}{cccc} 0 & 0 & 0 & -1  \\ 0 & 0 & 0 & 0 \\ 0 & 0 & 0 & 0 \\ 1 & 0 & 0 & 0   \end{array}\right) \quad
T_{34} = \left(\begin{array}{cccc} 0 & 0 & 0 & 0  \\ 0 & 0 & 0 & 0 \\ 0 & 0 & 0 & -1 \\ 0 & 0 & 1 & 0   \end{array}\right)
\]            
As they stand these generators do not form a closed algebra.  But we may combine them to form \emph{two} sets of generators, $\{i,j,k\}$ and $\{l,m,n\}$, each of which forms a closed algebra, as follows.   
\[
i =  T_{12} + T_{34} = \left(\begin{array}{cccc} 0 & -1 & 0 & 0  \\ 1 & 0 & 0 & 0 \\ 0 & 0 & 0 & -1 \\ 0 & 0 & 1 & 0 \end{array}\right)
\]
\[
j =  T_{13} - T_{24}  = \left(\begin{array}{cccc} 0 & 0 & -1 & 0  \\ 0 & 0 & 0 & 1 \\ 1 & 0 & 0 & 0 \\ 0 & -1 & 0 & 0 \end{array}\right)
\]
\[
k =  T_{23} + T_{14} = \left(\begin{array}{cccc} 0 & 0 & 0 & -1  \\ 0 & 0 & -1 & 0 \\ 0 & 1 & 0 & 0 \\ 1 & 0 & 0 & 0 \end{array}\right)  
\]
and 
\[
l = T_{12} - T_{34} = \left(\begin{array}{cccc} 0 & -1 & 0 & 0  \\ 1 & 0 & 0 & 0 \\ 0 & 0 & 0 & 1 \\ 0 & 0 & -1 & 0 \end{array}\right) 
\]
\[
m =  T_{13} + T_{24} = \left(\begin{array}{cccc} 0 & 0 & -1 & 0  \\ 0 & 0 & 0 & -1 \\ 1 & 0 & 0 & 0 \\ 0 & 1 & 0 & 0 \end{array}\right)  
\]
\[
n =  T_{23} - T_{14} = \left(\begin{array}{cccc} 0 & 0 & 0 & 1  \\ 0 & 0 & -1 & 0 \\ 0 & 1 & 0 & 0 \\ -1 & 0 & 0 & 0 \end{array}\right)   
\]

If we add the identity matrix $I$ to our list we have the following multiplication tables 

\begin{equation*}
\begin{array}{c|cccc}
	& I & i & j & k \\
	\hline 
	I & I & i & j & k \\ 
	i & i & -I & k & -j \\
	j & j & -k & -I  & i \\
	k & k & j & -i & -I  \\
\end{array} 
\qquad \qquad
\begin{array}{c|cccc}
	& I & l & m & n \\
	\hline 
	I & I & l & m & n \\ 
	l & l & -I & n & -m \\
	m & m & -n & -I  & l \\
	n & n & m & -l & -I \\
\end{array} 
\end{equation*}
Note that each element of the set $\{i,j,k\}$ commutes with each element of $\{l,m,n\}$, e.g. $il=li$ etc.  The left table above can be summarized as
\[
i^2 = j^2 = k^2 = ijk = -I,
\]
which is isomorphic to Hamilton's famous quaternion equation, with $I$ standing for $1$ and $\{i,j,k\}$ standing for Hamilton's triplet of imaginary numbers.  A similar quaternion equation holds for $\{l,m,n\}$.\footnote{Another representation of quaternions is found in the triplet $(-i\sigma_x, -i\sigma_y, -i\sigma_z)$ where the $\sigma$ are the Pauli spin matrices. Using the 2x2 representation of i from section 2, this representation is the same as $(n, m, -l)$.  For further discussion of the relationship between quaternions and spinors see Kronsbein (1967).}  If we restrict computation to actions involving $\{I,i,j,k\}$ without reference to the underlying vector space, we can simply replace $I$ with $1$ and treat our quartet as a quaternion.  However in the remainder of this section we will continue to utilize the symbol $I$.   

Consider now an arbitrary quaternion $R = wI+xi + yj + zk$.  This can always be written in the form
\begin{equation}\label{R}
R = r  \left \{ \cos \theta \, I + \sin \theta   ( u_x i + u_y j + u_z k ) \right \}, \quad r>0, \quad u_x^2 + u_y^2 + u_z^2 = 1    
\end{equation}
The quantity in the curly bracket can be re-written
\[
\cos \theta \, I + \sin \theta   ( u_x i + u_y j + u_z k  ) = \mathop {\lim } \limits_{N \to \infty} \left [ I+  \frac{\theta}{N} ( u_x i + u_y j + u_z k  ) \right ]^N,
\]
which is analogous to equation \eqref{generator} and can be proved in a similar manner. So we may say that $\{i,j,k\}$ are the generators of rotations in our four dimensional vector space.    

The mapping between quaternions and vectors in four dimensions can be demonstrated using the following basis vectors:
\[
\bm{I} = I \left(\begin{array}{c} 1 \\ 0 \\ 0 \\ 0 \end{array}\right ) = \left(\begin{array}{c} 1 \\ 0 \\ 0 \\ 0 \end{array}\right ) \\  
\]
\[
\bm{i} = i \left(\begin{array}{c} 1 \\ 0 \\ 0 \\ 0 \end{array}\right ) = \left(\begin{array}{c} 0 \\ 1 \\ 0 \\ 0 \end{array}\right ) \\
\]
\[
\bm{j} = j \left(\begin{array}{c} 1 \\ 0 \\ 0 \\ 0 \end{array}\right ) = \left(\begin{array}{c} 0 \\ 0 \\ 1 \\ 0 \end{array}\right ) \\
\]
\begin{equation} \label{Isomorph}
\bm{k} = k \left(\begin{array}{c} 1 \\ 0 \\ 0 \\ 0 \end{array}\right ) = \left(\begin{array}{c} 0 \\ 0 \\ 0 \\ 1 \end{array}\right )  
\end{equation}
\[
\bm {l} = l \left(\begin{array}{c} 1 \\ 0 \\ 0 \\ 0 \end{array}\right) = \left(\begin{array}{c} 0 \\ 1 \\ 0 \\ 0 \end{array}\right) \\
\]
\[
\bm {m} = m \left(\begin{array}{c} 1 \\ 0 \\ 0 \\ 0 \end{array}\right) = \left(\begin{array}{c} 0 \\ 0 \\ 1 \\ 0 \end{array}\right) \\
\]
\[
\bm {n} = n \left(\begin{array}{c} 1 \\ 0 \\ 0 \\ 0 \end{array}\right) = \left(\begin{array}{c} 0 \\ 0 \\ 0 \\ -1 \end{array}\right) 
\] 

Given this duality of four-dimensional vectors and quaternions one may consider $(w,x,y,z)$ as either representing a vector in four dimensions or as representing a rotation and a scaling operation.  This duality property was known to Hamilton and was emphasized by his successor Tait, who considered it a key property of quaternions (Silva and Martins, 2002).  It is in the sense of this duality that one may consider quaternions to be the natural extension of complex numbers.  

We now turn to the question of how a system of four-dimensional rotations can give rise to a system of three dimensional geometry.  Consider the transformation of a quaternion $\Phi$ by some other quaternion $Q$:
\begin{equation*}
\Psi   = Q  \Phi 
\end{equation*}
Now imagine another unit-quaternion $R$ being applied to both $ \Phi $ and $ \Psi $:
\begin{equation} \label{trans} 
\Phi^\prime  = R\Phi, \quad
\Psi^\prime  = R\Psi 
\end{equation}     
Since $R$ is of unit magnitude ($r=1$ in Equation \eqref{R}), it preserves the norms of $ \Phi $ and $ \Psi $.  We want to enquire as to the new relationship between $ \Phi^\prime $ and $ \Psi^\prime $,  That is, to find the new quaternion $Q^\prime$ such that
\begin{equation}\label{primed}
\Psi^\prime  = Q^\prime \Phi^\prime.
\end{equation}
Let $R = aI+bi+cj+dk$ where $a^2+b^2+c^2+d^2=1$, and define $R^\dagger \equiv aI-bi-cj-dk$.  The following property is readily verified:
\begin{equation*}
R^\dagger R= R R^\dagger = I
\end{equation*}   
It then follows from \eqref{trans} and \eqref{primed} that
\begin{equation*} 
Q^\prime = RQR^\dagger
\end {equation*}
which is a similarity transformation.          

Defining $Q=wI+xi+yj+zk$, $Q^\prime=w^\prime I+x^\prime i+y^\prime j+z^\prime k$, $\bm v = (x,y,z)$, $\bm v^\prime = (x^\prime ,y^\prime ,z^\prime )$ and using the definition of $R$ in Equation \eqref{R} with $r=1$, one may show that 
\begin{equation*} 
w^\prime = w 
\end{equation*}
and
\begin{equation*} 
\bm {v'} = \bm v_\| + \bm v_\bot \cos 2\theta + (\bm u \times \bm v_\bot ) \sin 2\theta, 
\end{equation*}
where $\bm v_\|=\bm u (\bm u \cdot \bm v)$ and $\bm v_\bot = \bm v - \bm u (\bm u \cdot \bm v)$.  This is none other than Equation \eqref{3} but with twice the angle!    

And so we see how the geometry of three dimensions is reclaimed.  The portion of the quaternion $xi+yj+zk$ (called a ``pure quaternion'' by Hamilton) is transformed under rotations exactly like an ordinary three-dimensional polar vector, while the ``scalar'' part of the quaternion $wI$ remains unchanged.  The factor of $2$ appearing with the angle $\theta$ is  consistent with the image of the parameter space hypothesized in the previous section.  The angle $\theta$ in Equation \eqref{R} can be interpreted as the line of latitude circling the four-dimensional hypersphere (of radius $r$), and the unit vector $(u_x,u_y,u_z)$ can be interpreted as the direction of longitude away from the north pole of the hypersphere.  Note that since the elements of $\{i,j,k\}$ commute with the elements of $\{l,m,n\}$, quaternions built out of $\{i,j,k\}$ are invariant with respect to similarity transformations based on $\{l,m,n\}$, and vice versa.  So there are actually two separate three-dimensional worlds within the four-dimensional system that we have constructed.  The relationship between these two three-dimensional worlds is considered next.   

A pure quaternion $xi+yj+zk$ behaves like a polar vector under rotation but it does not behave like a polar vector under reflection.  To illustrate, consider the special case of a quaternion that is the product of two pure quaternions $Q = (xi+yj+zk)(ui+vj+wk)$.  Consider the reflection of $xi+yj+zk$, $ui+vj+wk$ and $Q$ through a mirror oriented in the x-y plane.  If $xi+yj+zk$ acts like a polar vector, then upon reflection it should become $xi+yj-zk$.  Similarly for $ui+vj+wk$.  But the sign of the k$^{th}$ component of $Q^\prime = (xi+yj-zk)(ui+vj-wk)$ does \emph{not} become flipped, so our treatment of reflections is not consistent.  Silva and Martins (2002) avoid this inconsistency by treating pure quaternions as axial vectors.  Equation \eqref{Isomorph} suggests another way to define reflections for quaternions.  The unit vectors $\{\bm{i},\bm{j},\bm{k}\}$ form the basis of a right-handed coordinate system while the unit vectors $\{\bm{l},\bm{m},\bm{n}\}$ form the basis of a left-handed coordinate system (through reflection in the x-y plane).  By extension, the mirror image of $\{i,j,k\}$ should be $\{l,m,n\}$, which leads to a consistent transformation of $Q$.  And so we see that the two three-dimensional worlds discussed in the previous paragraph are located on opposite sides of the looking glass.

\section{Conclusion}
The purpose of this paper has been to develop an understanding of complex numbers and quaternions by focusing on the properties of the parameter-space of rotations and on the mapping between these parameter-spaces and Euclidean vector spaces.  In two dimensions, a complex number can represent a vector in the Argand plane, or it can represent a rotation.  In four dimensions a similar duality gives rise to quaternions.  One can reclaim three-dimensional geometry from quaternions by analyzing the effects of similarity tansformations.  The six generators of rotations in four dimensions decompose into two sets of three-dimensional generators, and these sets are related to each other by a parity transformation.                    

\emph{The author would like to thank E. Poisson, D. Staley and two anonymous referees for helpful suggestions on improving the manuscript.}


\begin{thebibliography}{9}

\bibitem{E1}
Bell, E.T.,
\emph{Men of Mathematics},
Simon and Schuster, 1937.

\bibitem{E2}
Bohm, David,
Space-Time Geometry as an Abstraction from Spinor Ordering, in \emph{Perspectives in Quantum Theory}, edited by Wolfgang Yourgrau and Alwyn van der Merwe, M.I.T. Press, 1971.

\bibitem{E3}
Coddens, G., Spinor approach to the rotation and reflection groups, \emph{European Journal of Physics} $\bm{23}$, 2002, 549-564.

\bibitem{E4}
Crowe, Michael,
\emph{A History of Vector Analysis},
University of Notre Dame Press, 1967.

\bibitem{E5}
Frescura, F., Hiley, B., Geometric interpretation of the Pauli spinor, \emph{American Journal of Physics} $\bm{49}$, February 1981, 152-157.

\bibitem{E6}
Girard, P., The quaternions group and modern physics, \emph{European Journal of Physics} $\bm{5}$, 1984, 25-32.

\bibitem{E7}
Joshi, A. W., \emph{Elements of Group Theory for Physicists, Third Ed.}, John Wiley and Sons, 1982.

\bibitem{E8}
Kronsbein, John, Kinematics - Quaternions - Spinors - and Pauli's Spin Matrices, \emph{American Journal of Physics} $\bm{35}$, April 1967, 335-342.

\bibitem{E9}
Kuipers, Jack,
\emph{Quaternions and Rotation Sequences},
Princeton University Press, 1999.

\bibitem{E10}
Misner, C., Thorne, K. and Wheeler, J., \emph{Gravitation},
W. H. Freeman and Company, 1973.

\bibitem{E11}
Penrose, R. and Rindler, W., \emph{Spinors and Space-time},
Cambridge University Press, 1986.

\bibitem{E12}
Silva, C. and Martins, R., Polar and axial vectors versus quaternions, \emph{American Journal of Physics} $\bm{70}$, September 2002, 958-963.

\bibitem{E13}
Silverman, M., The curious problem of spinor rotation, \emph{European Journal of Physics} $\bm{1}$, 1980, 116-122.

\bibitem{E14}
Stephenson, Reginald J., Development of Vector Analysis from Quaternions, \emph{American Journal of Physics} $\bm{34}$, March 1966, 194-201.

\bibitem{E15}
van der Waerden, B.L., Hamilton's discovery of quaternions, \emph{Mathematics Magazine} $\bm{49}$, November 1976, 227-234.

\end{thebibliography}
\end{document}